\documentclass[preprint,showpacs,amsfonts,amsmath,amssymb]{revtex4}
\usepackage{graphicx}
\usepackage{dcolumn}
\usepackage{bm}
\usepackage{showlabels}

\begin{document}
\title{ Timelike Geodesic Motion in Ho$\breve{r}$ava-Lifshitz
Spacetime }
\author{Juhua Chen$^{1,2}$} \email{jhchen@hunnu.edu.cn}
\author{Yongjiu Wang$^{1}$}
\affiliation{College of Physics and Information Science, Hunan
Normal University, Changsha, Hunan 410081, P. R. China
\\Department of Physics \& Astronomy,
      University of Missouri, Columbia, MO 65211, USA}
\begin{abstract}
Recently Ho$\breve{r}$ava proposed a non-relativistic renormalisable
theory of gravitation. When restricted to satisfy the condition of
detailed balance, this theory is intimately related to topologically
massive gravity in three dimensions, and the geometry of the Cotton
tensor. At long distances, this theory is expected to flow to the
relativistic value $\lambda = 1$, and could therefore serve as a
possible candidate for a UV completion of Einstein general
relativity or an infrared modification thereof. In this paper under
allowing the lapse function to depend on the spatial coordinates
$x^i$ as well as $t$, we obtain the spherically symmetric solutions.
And then by analyzing the behavior of the effective potential for
the particle, we investigate the timelike geodesic motion of
particle in the Ho$\breve{r}$ava-Lifshitz spacetime. We find that
the nonradial particle  falls from a finite distance to the center
along the timelike geodesics when its energy is in an appropriate
range. However, we find that it is complexity for radial particle
along the timelike geodesics. There are follow different cases due
to the energy of radial particle: 1) When the energy of radial
particle is higher than a critical value $E_{C}$, the particle will
fall from infinity to the singularity directly; 2) When the energy
of radial particle equals to the critical value $E_{C}$, the
particle orbit is unstable at $r=r_{C}$, i.e. the particle will
escape from $r=r_{C}$ to the infinity or to the singularity, which
is determined by the initial conditions of the particle; 3) When the
energy of radial particle is in a proper range, the particle will
rebound to the infinity or plunge to the singularity from a infinite
distance, which is also determined by the initial conditions of the
particle.

\pacs{04.70.-s, 95.30.Sf, 97.60.Lf}

\end{abstract}
\maketitle
\section{Introduction}
P. Ho$\breve{r}$ava suggested a renormalizable four-dimensional
theory of gravity, which admits the Lifshitz scale-invariance in
time and space \cite{hor1,hor2}. The theory is with a broken Lorentz
symmetry at short distances, while at large distances, higher
derivative terms do not contribute and the theory runs to the
standard GR, if a coupling, which controls the contribution of trace
of the extrinsic curvature has a particular value ($\lambda=1$). In
this specific limit, post- Newtonian corrections of the
Ho$\breve{r}$ava Gravity coincide with those of the standard GR.

Up to now there are several versions of Ho$\breve{r}$ava gravity,
which are classified according to whether or not the detailed
balance and the projectability conditions are imposed. Since
Ho$\breve{r}$ava theory is modelled after a scalar field model
studied by Lifshitz \cite{Lifshitz}, the theory is called
Ho$\breve{r}$ava-Lifshitz (HL) gravity. So far most of the work on
the HL theory has abandoned the projectability condition but
maintained detailed balance
\cite{Wang1,Wang2,Wang3,Wang4,Wang5,Wang6,Wang7,Wang8,Wang9,Wang10,Wang11,Wang12,Wang13,Wang14,Wang15,Wang16,Wang17,Wang18}.
One of the main reasons is that the resulting theory is much simpler
to deal with, giving local rather than global energy constraints.
Specific solutions of this simplest version of Ho$\breve{r}$ava-
Lifshitz gravity have recently been analyzed
\cite{Kehagias1,Kehagias2,Kehagias3,Kehagias4,Kehagias5,Kehagias6,Kehagias7}.
Homogeneous vacuum solutions with gravitational waves were studied
in Ref.\cite{Takahashi}, and black hole solutions were analyzed in
Ref.\cite{Lu1,Lu2,Lu3,Lu4,Lu5}. G. Calcagni et al
\cite{Calcagni1,Calcagni2,Calcagni3,Calcagni4,Calcagni5} gave
cosmological solutions with a Lifshitz scalar mater which is the
analogs of the Friedmann equations in Ho$\breve{r}$ava-Lifshitz
gravity include a term which scales as dark radiation and
contributes a negative term to the energy density. Thus, it is
possible in principle to obtain a nonsingular cosmological evolution
with the Big Bang of Standard and Inflationary Cosmology replaced by
a bounce \cite{Brandenberger1,Brandenberger2}. In last months some
author focused on investigating the various properties and
consequences of the Ho$\breve{r}$ava-Lifshitz gravity. Konoplya
considered the potentially observable properties of black holes in
the Ho$\breve{r}$ava-Lifshitz gravity with Minkowski vacuum: the
gravitational lensing and quasinormal modes
\cite{Konoplya1,Konoplya2}. Nishioka \cite{Nishioka} derived the
detailed balance condition as a solution to the Hamilton- Jacobi
equation in the Ho$\breve{r}$ava-Lifshitz gravity, which proposes
the existence of the d-dimensional quantum field theory with the
effective action on the future boundary of the (d + 1)-dimensional
Ho$\breve{r}$ava-Lifshitz gravity from the viewpoint of the
holographic renormalization group. Myung \cite{Myung1, Myung2,
Myung3} studied the thermodynamics of black holes in the deformed
Ho$\breve{r}$ava-Lifshitz gravity. By using the canonical
Hamiltonian method, Rong-Gen Cai \cite{Cai1,Cai2,Cai3,Cai4,Cai5}
obtained the mass and entropy of the black holes with general
dynamical coupling constant $\lambda$ in Ho$\breve{r}$ava-Lifshitz
Gravity.

Timelike geodesic of  particles in a given space-time is very
interesting topics in general relativity. By using the method of an
effective potential, Jaklitsch {\sl{et al}} \cite{Jaklitsch}
investigated the time-like geodesic structure with a positive
cosmological constant and Cruz {\sl{et al}} \cite{Cruz} studied the
geodesic structure of the Schwarzschild Anti-de Sitter black hole.
The analysis of the effective potential for null geodesics in
Reissner-Nordstr\"{o}m-de Sitter and Kerr-de Sitter space-time was
carried out in Ref.\cite{Stuchlik}. Podolsky \cite{Podolsky}
investigated all possible geodesic motions in the extreme
Schwarzschild-de Sitter space-time. Chen and Wang
\cite{chen1,chen2,chen3,chen4,chen5,chen6,chen7} have investigated
the orbital dynamics of the test particle in several gravitational
fields with an electric dipole and a mass quadrupole, and in the
extreme Reissner-Nordstr\"{o}m black hole space-time. The exact
solutions in the closed analytic form for the geodesic motion in the
Kottler space-time were considered by Kraniotis \cite{Kraniotis1}.
The exact solutions of the time-like geodesics were used to
investigate the perihelion precession of the planet Mercury. By
solving the Hamilton-Jacobi partial differential equation, Kraniotis
and Whitehouse \cite{Kraniotis2} investigated the geodesic motion of
the massive particle in the Kerr and Kerr-(anti)de Sitter
gravitational field. In this paper we plan to extend Jaklitsch's
effective potential method to concentrate on the time-like geodesic
motion in H$\breve{o}$rava-Lifshitz Spacetime.

\section{Timelike geodesic equation and effective potential}
Recently a new four-dimensional non relativistic renormalizable
theory of gravity  was proposed by Ho\v{r}ava \cite{hor2}. It is
believed that this theory is a  UV completion for the Einstein
theory of gravitation. We use the ADM decomposition method
\cite{Lu1,Lu2,Lu3,Lu4,Lu5} to find spherically symmetric solutions.
We start from the four-dimensional metric written in the ADM
formalism, Let us consider the ADM decomposition of the metric in
standard GR,
\begin{eqnarray}
 ds^2=-N^2 dt^2+g_{ij}\left(dx^i+N^i dt\right)\left(dx^j+N^j
dt\right),
\end{eqnarray}
where the lapse, shift and 3-metric $N, N^i$ and $g_{ij}$ are all
functions of $t$ and $x^i$. The lapse function N is viewed as a
gauge field for time reparameterisations, and it is effectively
restricted to depend only on $t$, but not the spatial coordinates
$x^i$. A closer parallel with general relativity is achieved if this
projectability restriction is relaxed. The action for the fields of
the theory is
\begin{eqnarray}\label{action1}
 S &= & \int dt d^3 x
\sqrt{g}N\left\{\frac{2}{\kappa^2}(K_{ij}K^{ij}-\lambda
K^2)-\frac{\kappa^2}{2w^4}C_{ij}C^{ij}+\frac{\kappa^2
\mu}{2w^2}\epsilon^{ijk} R^{(3)}_{i\ell} \nabla_{j}R^{(3)\ell}{}_k
\right.
\nonumber \\
&&\left. -\frac{\kappa^2\mu^2}{8} R^{(3)}_{ij}
R^{(3)ij}+\frac{\kappa^2 \mu^2}{8(1-3\lambda)}
\left(\frac{1-4\lambda}{4}(R^{(3)})^2+\Lambda_W R^{(3)}-3
\Lambda_W^2\right)+\mu^4 R^{(3)}\right\}.
\end{eqnarray}

We should note that
\begin{eqnarray}
K_{ij}=\frac{1}{2N}\left(\dot{g}_{ij}-\nabla_i
N_j-\nabla_jN_i\right)\ ,
\end{eqnarray}
is the second fundamental form, and
\begin{eqnarray}
 C^{ij}=\epsilon^{ik\ell}\nabla_k
\left(R^{(3)j}{}_\ell-\frac{1}{4}R^{(3)} \delta^j_\ell\right)\ ,
\end{eqnarray}
is the Cotton tensor.  $\kappa,\lambda,w$ are dimensionless coupling
constants, whereas $\mu,\Lambda_W$ are dimensionfull of mass
dimensions $[\mu]=1,[\Lambda_W]=2$. The action (\ref{action1}) is
the action in Ref.\cite{hor2} where we have added the last term,
which represents a soft violation of the detailed balance condition.

We will now consider the limit of this theory, i.e.
$\Lambda_{W}\rightarrow 0$. In this particular limit, the theory
turns out to be
\begin{eqnarray}
 S&=&\int dt d^3 x
\sqrt{g}N\left\{\frac{2}{\kappa^2}\left(K_{ij}K^{ij}-\lambda
K^2\right)-\frac{\kappa^2}{2w^4}C_{ij}C^{ij}  +\frac{\kappa^2
\mu}{2w^2}\epsilon^{ijk} R^{(3)}_{i\ell} \nabla_{j}R^{(3)\ell}{}_k
\right.
\nonumber \\
&&\left. -\frac{\kappa^2\mu^2}{8} R^{(3)}_{ij} R^{(3)ij}
+\frac{\kappa^2 \mu^2}{8(1-3\lambda)}
\frac{1-4\lambda}{4}(R^{(3)})^2+\mu^4 R^{(3)}\right\}.
\end{eqnarray}
For the $\lambda=1(\omega=16\mu^2/\kappa^2)$ case, we get the
static, spherically symmetric metric for asymptotically flat
space-time
\begin{eqnarray} \label{metric}
ds^2=-f(r)dt^2+f(r)^{-1}dr^2 +r^2(d\theta^2+sin^2\theta d\varphi^2),
\end{eqnarray}
where the lapse function is given by
\begin{eqnarray}
f(r)=1+\omega r^2-\sqrt{r(\omega^2r^3+4\omega M)},
\end{eqnarray}
where $M$ is an integration constant, with dimension [M]=-1. For
$r\gg (M/\omega)^{1/3}$, we can get the usual behavior of
Schwarzschild black hole

\begin{eqnarray}
f(r)\approx 1+\frac{2M}{r}+\Game (r^{-4}),
\end{eqnarray}

The lapse function vanished at the zeros of the equation
\begin{eqnarray}
1+\omega r^2-\sqrt{r(\omega^2r^3+4\omega M)}=0,
\end{eqnarray}
there are two event horizons at
\begin{eqnarray}
r_{\pm}=M\left(1\pm \sqrt{1-\frac{1}{2\omega M^2}}\right).
\end{eqnarray}
Avoiding naked singularity at the origin implies $\omega M^2\geq
1/2$. When $\omega M^2\gg 1$ means the regime of the convetional
General Relativity, the outer horizon approaches
\begin{eqnarray}
r_{+}\approx 2M-\frac{1}{4\omega M^2}+...
\end{eqnarray}

The outer event horizon is lower than the usual Schwarzschild event
horizon, $r_{+ Sch}=2M$, and the inner one will approach the
singularity $r_{-}\rightarrow 0$.

In order to investigate the time-like geodesics of particles in the
space-time which is described by (\ref{metric}), we solve the
Euler-Lagrange equation for the variation problem associated with
this metric. The corresponding Lagrangian is
\begin{eqnarray}\label{Lagrangian}
\L=-f(r)\dot{t}^2+f(r)^{-1}\dot{r}^2 +r^2(\dot{\theta}^2+sin^2\theta
\dot{\varphi}^2),
\end{eqnarray}
where the dots represent the derivative with respect to the affine
parameter $\tau$, along the geodesics. The equations of motion are
\begin{eqnarray}
\dot{\Pi}_q-\frac{\partial \L}{\partial q}=0,
\end{eqnarray}
where $\Pi_q=\frac{\partial \L}{\partial \dot{q}}$ is the momentum
conjugate to coordinate $q$. For the Lagrangian is independent of
$(t,\varphi)$, the corresponding conjugate momentum is conserved,
thus
\begin{eqnarray}\label{momenta1}
\Pi_t=-(1+\omega r^2-\sqrt{r(\omega^2r^3+4\omega M)})\dot{t}=-E,
\end{eqnarray}
\begin{eqnarray}
\Pi_\varphi=r^2sin^2\theta \dot{\varphi}=L.
\end{eqnarray}
Choosing the conditions $\theta=\pi/2$ and $\dot{\theta}=0$, then
the above equation will be simplified into
\begin{eqnarray}\label{momenta2}
\Pi_\varphi=r^2 \dot{\varphi}=L.
\end{eqnarray}
Substituting Eqs(\ref{momenta1}, \ref{momenta2}) into
Eq(\ref{Lagrangian}), we obtain
\begin{eqnarray}
2\L=h&=&\frac{E^2}{1+\omega r^2-\sqrt{r(\omega^2r^3+4\omega M)}}
\nonumber \\&-&\frac{\dot{r}^2}{1+\omega
r^2-\sqrt{r(\omega^2r^3+4\omega M)}}-\frac{L^2}{r^2}.
\end{eqnarray}
Now we solve the above equation for $\dot r^{2}$ in order to
obtain the radial equation, which allows us to characterize
possible moments of test particles without an explicit solution of
the equation of motion in an invariant plane
\begin{eqnarray}\label{radial}
\dot r^{2}= E^{2}-\left(1+\omega r^2-\sqrt{r(\omega^2r^3+4\omega
M)}\right)\left(h+ \frac{L^{2}}{r^{2}}\right).
\end{eqnarray}
We can rewrite above equation as a one-dimensional one
\begin{eqnarray}
\dot r^{2}= E^{2}-V^{2}_{eff},
\end{eqnarray}

\section{Timelike geodesics}
For the time-like geodesics, $h=1$, and  the effective potential
turns into
\begin{eqnarray}
V^{2}_{eff}=\left(1+\omega r^2-\sqrt{r(\omega^2r^3+4\omega
M)}\right)\left(1+ \frac{L^{2}}{r^{2}}\right).
\end{eqnarray}
From  effective potential for particle along the time-like geodesic,
we investigate the radial motion equation (\ref{radial}) for two
cases: non-radial and radial particles in the following subsections,
respectively.

\subsection{Timelike geodesics of radial particles}
The radial geodesic corresponds to the motion of the particle
without angular momentum $L=0$, which means the particles fall from,
in a finite distance, to the center. So the effective potential
$V^2_{eff}$  has the following form
\begin{eqnarray}
V^{2}_{eff}=1+\omega r^2-\sqrt{r(\omega^2r^3+4\omega M)},
\end{eqnarray}

\begin{figure}[htbp]
\begin{center}
\includegraphics{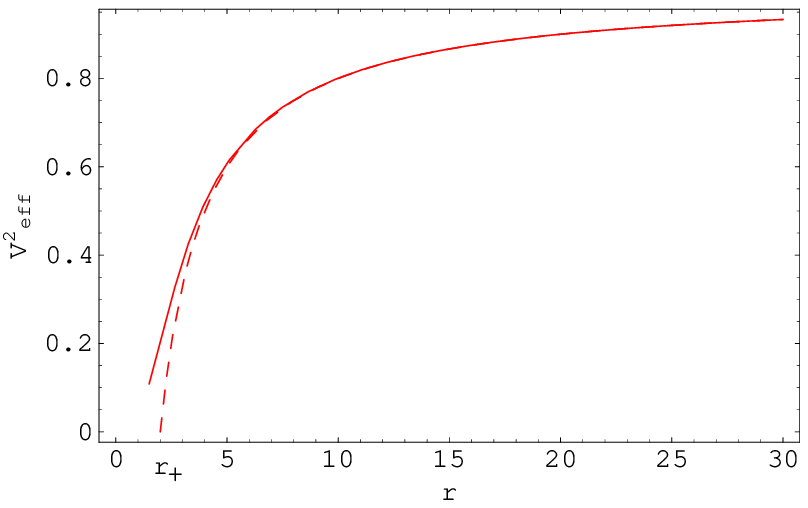}
\end{center}
\caption{The effective potentials $V^{2}_{eff}$ for radial particles
vs $r$ with the parameters $\omega=1, L=0$ and $M=1$. The dashed
line corresponds to the effective potential in the Schwarzschild
spacetime with the same parameter values. }
\begin{center}
\includegraphics{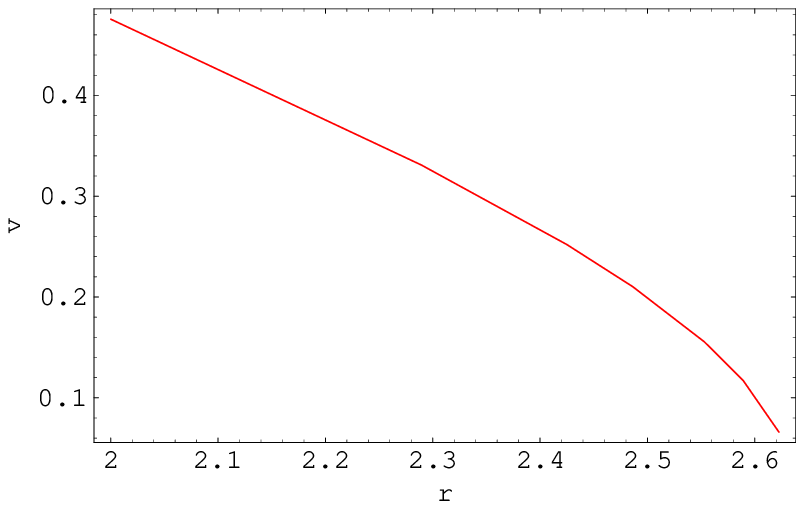}
\end{center}
\caption{The radial motion  of radial particles in ($v$ vs $r$)
phase space
 with the parameters $E=0.5, \omega=1$ and $M=1$.}
\end{figure}
and the corresponding radial motion equation takes the following
form:
\begin{eqnarray}
\dot r^{2}= E^{2}-\left(1+\omega r^2-\sqrt{r(\omega^2r^3+4\omega
M)}\right).
\end{eqnarray}

In order to investigate the properties of this effective potential,
 we simulate it in Fig.1. From this figure we can find that particles always plunge
into the horizon from an upper distance which is determined by the
motion energy $E$ of particles. At the same time we numerically
perform the evolution of the radial particles in (v-r) phase space
in Fig.2, we see that the velocity of particles, with a constant
energy $E$, will tend to zero when the particles reach a finite
distance, then radial particles will turn back and fall to the
center.

\subsection{ Timelike geodesics non-radial particles}

\begin{figure}[htbp]
\begin{center}
\includegraphics{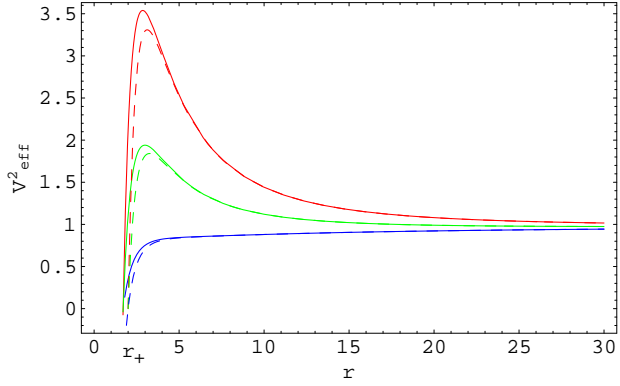}
\end{center}
\caption{The effective potentials $V^{2}_{eff}$ for non-radial
particles vs $r$ at different values of $L^2: 80M^2,40M^2$ and
$10M^2$(from top to bottom) with $\omega=1$ and $h=M=1$. The dashed
lines correspond to the effective potentials in the Schwarzschild
spacetime with the same parameter values.}
\end{figure}
\begin{figure}[htbp]
\begin{center}
\includegraphics{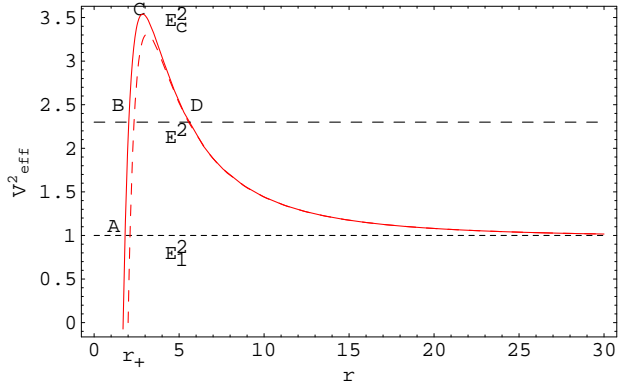}
\end{center}
\caption{The effective potentials $V^{2}_{eff}$ for non-radial
particles vs $r$ at different values of $L^2=80M^2, \omega=1$ and
$h=M=1$. The dashed line corresponds to the effective potential in
the Schwarzschild spacetime with the same parameter values.}
\end{figure}
\begin{eqnarray}\label{motion}
\dot r^{2}= E^{2}-\left(1+\omega r^2-\sqrt{r(\omega^2r^3+4\omega
M)}\right)\left(1+ \frac{L^{2}}{r^{2}}\right),
\end{eqnarray}
where $V^{2}_{eff}$ is defined as an effective potential and
expressed as
\begin{eqnarray}
V^{2}_{eff}=\left(1+\omega r^2-\sqrt{r(\omega^2r^3+4\omega
M)}\right)\left(1+ \frac{L^{2}}{r^{2}}\right).
\end{eqnarray}

We have plotted the effective potential for non-radial particles
with different values of $L^2: 80M^2,40M^2$ and $10M^2$(from top to
bottom) with $\omega=1$ and $h=M=1$ in Fig.3 as an example. From the
motion equation (\ref{motion}) and Fig.4,  we can divided the
time-like geodesics for non-radial particles into several cases
according to different values of $E$ for the non-radial particles as
follows:

(I) when $E^2>E^2_{C}$, the non-radial particle with enough energy
will directly fall from infinity to the singularity.

(II) when $E^2=E^2_{C}$, the orbit of the non-radial particle is
unstable at $r=r_{C}$. The non-radial particle in this kind of orbit
will escape from $r=r_{C}$ to the infinity, or to the singularity,
which is determined by the initial conditions of the particle.

(III) when $E^2_{1}<E^2<E^2_{C}$, the non-radial particle in this
kind of orbit will rebound from $r=r_{D}$ to the infinity, or to the
singularity from $r=r_{B}$, which is also determined by the initial
condition of the particle.

(IV) when $E^2<E^2_{1}$, the non-radial particle will plunge
 into the singularity directly.

\section{conclusions}
In this paper we have investigated timelike geodesic motion for
radial and non-radial particles in Ho$\breve{r}$ava-Lifshitz
spacetime. by analyzing the behavior of the effective potential for
the particle, we investigate the timelike geodesic motion of
particle in the Ho$\breve{r}$ava-Lifshitz spacetime. We find that
the nonradial particle  falls from a finite distance to the center
along the timelike geodesics when its energy is in an appropriate
range. However, we find that it is complexity for radial particle
along the timelike geodesics. There are follow different cases due
to the energy of radial particle: When the energy of radial particle
is higher than a critical value $E_{C}$, the particle will fall from
infinity to the singularity directly; When the energy of radial
particle equals to the critical value $E_{C}$, the particle orbit is
unstable at $r=r_{C}$, i.e. the particle will escape from $r=r_{C}$
to the infinity or to the singularity, which is determined by the
initial conditions of the particle; When the energy of radial
particle is in a proper range, the particle will rebound to the
infinity or plunge to the singularity from a infinite distance,
which is also determined by the initial conditions of the particle.
At the same time we have compared all cases of timelike geodesic
motion for radial and non-radial particles in the
Ho$\breve{r}$ava-Lifshitz spacetime  with those in the Schwarzschild
spacetime. We found that properties of Ho$\breve{r}$ava-Lifshitz
spacetime in large scale will run to the standard GR spacetime which
is suggested by Ho$\breve{r}$ava.

\section{Acknowledgments}
J.H. Chen is supported by the National Natural Science Foundation of
China under Grant No.10873004, the Scientific Research Fund of Hunan
Provincial Education Department under Grant No. 08B051, program for
excellent talents in Hunan Normal University and the State Key
Development Program for Basic Research Program of China under Grant
No. 2010CB832800.


\end{document}